\documentclass{PoS}
\usepackage{bbm} 
\usepackage{graphicx}
\usepackage{amsmath}
\usepackage{amssymb}
\usepackage{url}

\title{The Lefschetz thimble and the sign problem}

\ShortTitle{The Lefschetz thimble and the sign problem}

\author{\speaker{Luigi Scorzato}\thanks{Present address: Accenture AG -- Big Data Analytics ASG. 20, Rue de
    Pr{\'e}-Bois CH-1215 Gen{\`e}ve.}\\ E-mail: \email{luigi@scorzato.it}}

\abstract{In this talk I review the proposal to formulate quantum field theories (QFTs) on a Lefschetz thimble,
  which was put forward to enable Monte Carlo simulations of lattice QFTs affected by a sign problem. First I will
  review the theoretical justification of the approach, and comment on some open issues. Then, I will review the
  algorithms that have been proposed and are being tested to represent and simulate a lattice QFT on a Lefschetz
  thimble. In particular, I will review the lessons from the very first models of QFTs that have been studied with
  this approach.}

\FullConference{The 33rd International Symposium on Lattice Field Theory\\
		14 -18 July 2015\\
		Kobe International Conference Center, Kobe, Japan*}

\begin{document}

\section{Introduction}

Many important physical systems are characterized by complex actions, when formulated in terms of a path integral.
But, if the action $S$ is not real, then the real part of $e^{-S}$ is not positive semi-definite and it cannot be
interpreted as a probability distribution.  In these cases, Monte Carlo calculations are not applicable directly.
This is the so called {\em sign problem}.  Many techniques have been proposed to overcome this problem, with
important partial successes, but the sign problem is still unsolved for a variety of important physical systems and
parameter values.  In this context, any new idea that could improve our chances to simulate any of these models on
larger lattices than are feasible today is extremely valuable.

In these proceedings, I review the approach based on Lefschetz thimbles (LT), which was introduced three years ago
\cite{Cristoforetti:2012su,Cristoforetti:2013wha,Fujii:2013sra} and it is now being further developed and used by a
growing number of groups.  This review is meant to be both an introduction for those who are new to this approach,
and also a status update for those who are already familiar with it.

To introduce this approach it may be convenient to start from the observation that strongly oscillating, low
dimensional integrals are typically treated very effectively with the saddle point integration method. It is
therefore natural to ask whether some of those ideas can be useful to deal with strongly oscillating {\em
  functional} integrals.  To explore this, consider a classic example: the Airy function, which is defined as the
following symmetric integral over the real axis:
\begin{equation}
\label{eq:Airy}
{\rm Ai}(x) = \frac{1}{2\pi} \int_{\infty}^{\infty} e^{i(\frac{t^3}{3}+xt)} dt.
\end{equation}

\begin{figure}[ht]
\centering
\includegraphics[width=70mm]{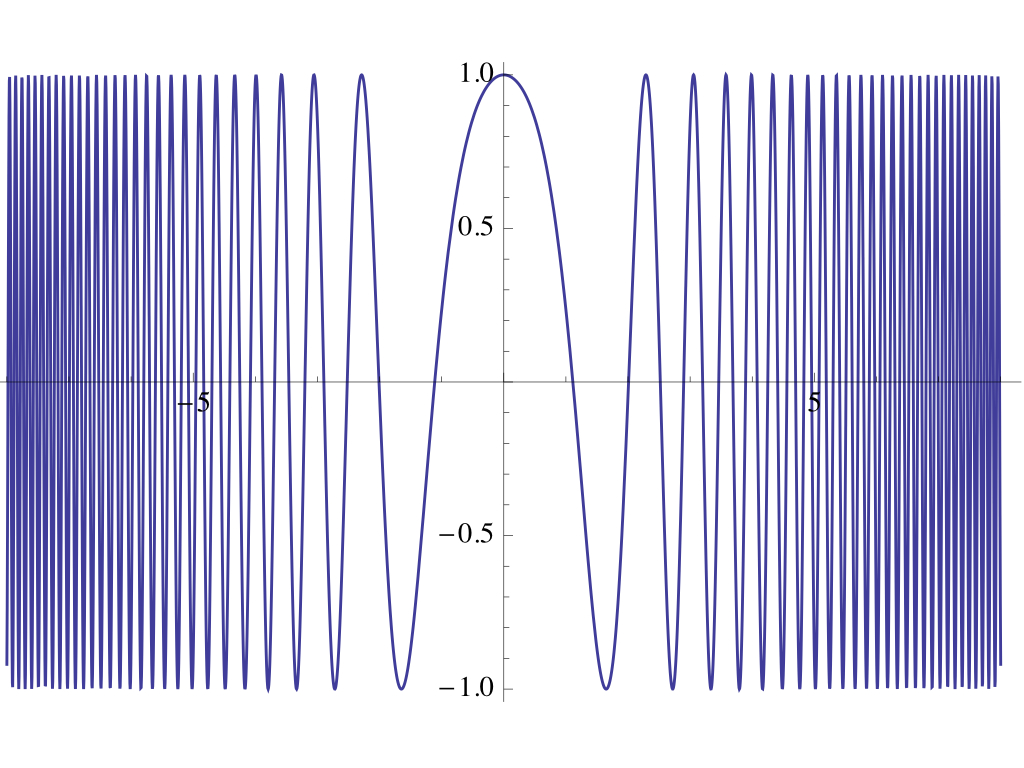}
\includegraphics[width=70mm]{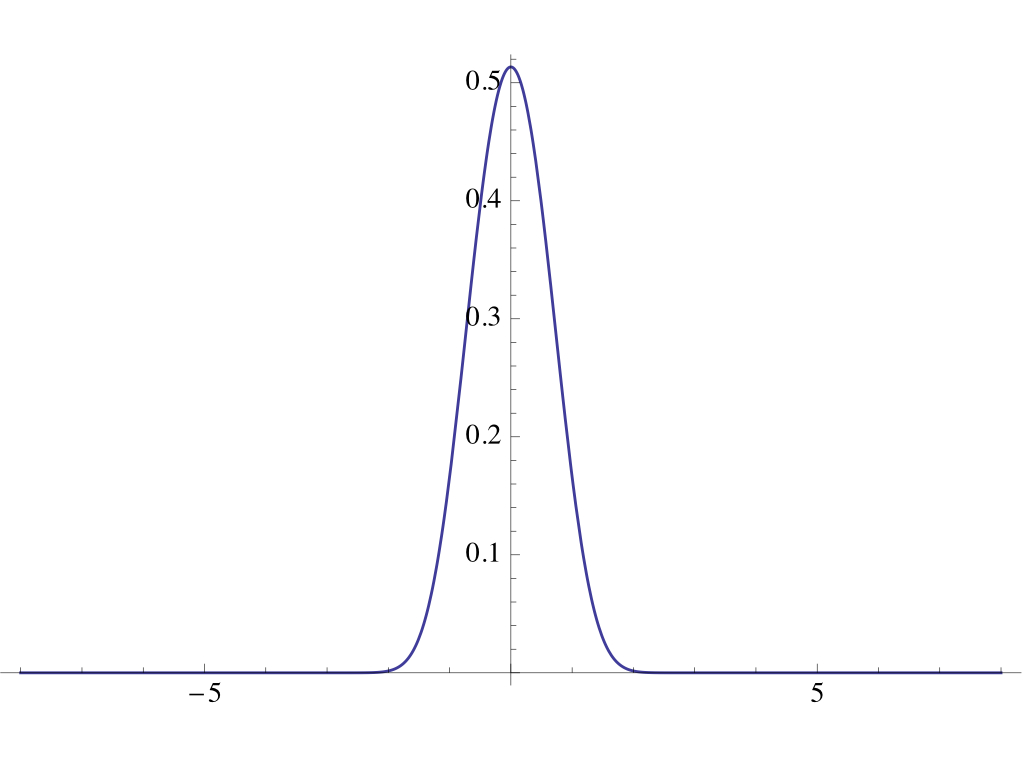}
\caption{Left panel: (real part of the) integrand defining the Airy function on the real axis
  (Re$[e^{i(\frac{t^3}{3}+xt)}]$). Right panel: the same integrand function along the curves of steepest
  descent. \label{fig:airy_curves}}
\end{figure}

\begin{figure}[ht]
\centering
\includegraphics[width=140mm]{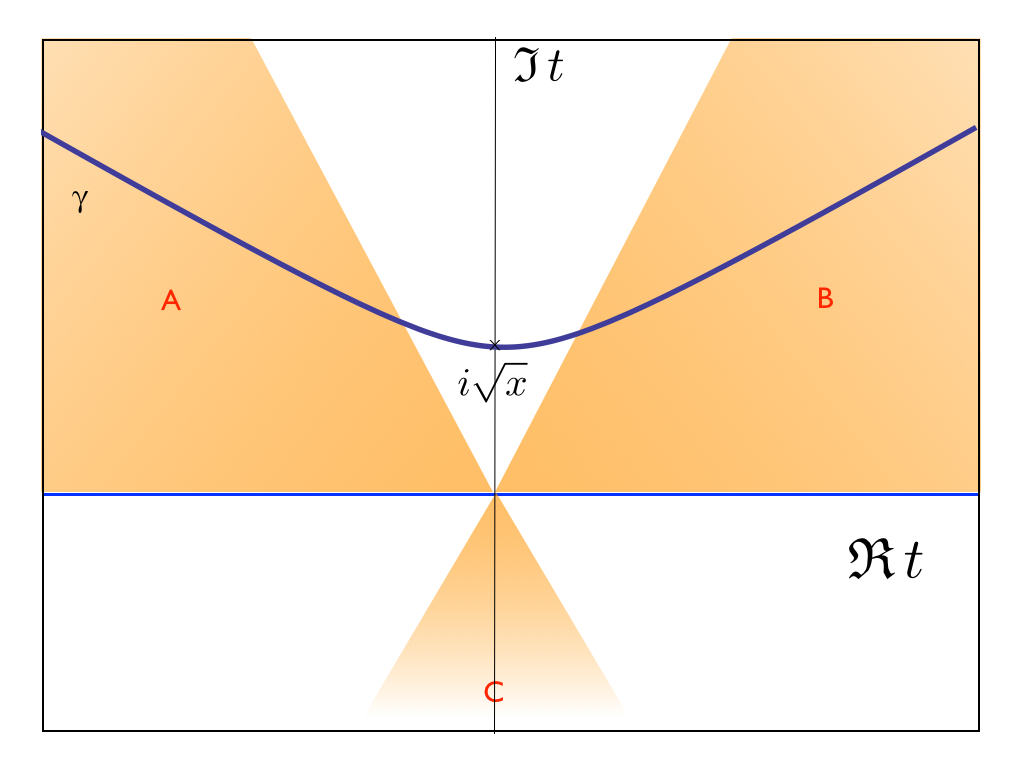}
\caption{In yellow are the regions of asymptotic convergence of the Airy function. In blue is the curve of steepest
  descent going through the stationary point $i\sqrt{x}$. \label{fig:airy}}
\end{figure}

The integrand is strongly oscillating on the real axis (Fig.~\ref{fig:airy_curves}, left), which makes a direct
numerical evaluation infeasible.  However, we know that we are free to deform the integration path in the complex
plane, as long as the new path belongs to the original relative homology class.  In this case, the original
relative homology class connects regions $A$ and $B$ at infinity (Fig.~\ref{fig:airy}), where:
\[
A:= \{t \in \mathbb{Z} \; | \; \frac{2}{3}\pi \, < \, \arg{t} < \pi \}
B:= \{t \in \mathbb{Z} \; | \; 0 \, < \, \arg{t} < \frac{1}{3}\pi \}.
\]

If we take this option, we also know that the optimal way to do it is to locate the stationary points of the
complexified function, and follow the paths of steepest descent ($\gamma(t)$), along which the integrand converges
optimally fast (Fig.~\ref{fig:airy_curves}, right), and the phase of the integrand is even constant, so that it can
be factorised out of the integral:
\begin{equation}
\frac{1}{2\pi} \int_{\gamma} e^{i(\frac{z^3}{3}+xz)}dz =
\frac{1}{2\pi} e^{i\phi} \int_{\gamma} e^{\mbox{Re}[\frac{z^3}{3}+xz]}dz.
\label{eq:statph}
\end{equation}
(Note that the tangent $\gamma'$ of the path $\gamma$ is not constant, which still leaves a nontrivial phase in the
measure of the integral on the rhs of Eq.~(\ref{eq:statph}), but it is expected to change smoothly. I will come
back to this in Sec.~\ref{sec:resph}.)

Saddle point integration method is a classic elementary tool that works very well for many low dimensional
oscillating integrals.  It is usually combined with an asymptotic series expansion around the stationary point.
But in our case, this would correspond to some version of perturbation theory, which is not what we want.  However
the idea of deforming the path is independent of the series expansion.  And a path where the phase is stationary
and the important contributions are more localized is very attractive from the point of view of the sign problem.
This naturally leads to the idea of performing a fully non-perturbative Monte Carlo integral along the curves of
steepest descent (SD). 

If we want to pursue this idea further, we need to understand what happens in higher dimensions.  The typical
integral that we want to consider has the form:
\[
{\cal I} = \int_{\mathbb{R}^n} dx^n \, g(x) \, e^{f(x)}.
\]
The generalization of the path of SD in $\mathbb{C}$ are then called {\em Lefschetz thimbles} (LT).  

For each stationary point $\varphi_{\sigma}$ of the complexified function $f(z)$, $z\in\mathbb{C}^n$, and for a
generic choice of its parameters, the LT ${\cal J}_{\sigma}$ is defined as the union of all the curves of SD that
fall in $\varphi_{\sigma}$ when $\tau\rightarrow\infty$, where:
\begin{equation}
\label{eq:SD}
\frac{dz_j}{d\tau} = - \frac{\partial\overline{f(z)}}{\partial\overline{z_j}}.
\end{equation}

Under suitable conditions on $f(z)$ and $g(z)$ \cite{Pham1983,Witten:2010cx,Witten:2010zr}\footnote{These include
  asymptotic conditions of convergence, which are typically fulfilled in physical applications, and --- importantly
  --- a generic choice of the parameters, as defined there.}, Morse/Picard-Lefschetz theory tells us that the
thimbles ${\cal J}_{\sigma}$ are smooth manifolds of real dimension $n$ immersed in $\mathbb{C}^n$ (i.e. the same
dimension as the original integration domain), and for each domain ${\cal C}$ where the integral converges, we
have:
\[
{\cal I} = \int_{\mathbb{R}^n} dx^n \, g(x) \, e^{f(x)} = 
\sum_{\sigma} n_{\sigma} \int_{{\cal J}_{\sigma}} dz^n \,
g(z) \, e^{f(z)}.
\]
In other words, the thimbles represent a basis of the relative homology group for the integral above, with some
integer coefficients $n_{\sigma}$.  For example, the integrand defining the Airy function has two stationary points
$\varphi_{\pm} = \pm \sqrt{-x}$ for $x\neq0$.  When $x<0$, these two stationary points define two disjoint thimbles
through them, and every integration path for the Airy function can be expressed in terms of them.  When $x>0$ (as in
Fig.~\ref{fig:airy}), there is also a path that connects the two stationary points, but an arbitrarily small
perturbation of $f$ can always restore the generic thimbles structure.

Can we use the thimble basis to compute the path integral of a quantum field theory (QFT)?  In principle yes: we
could write an expectation value as:
\[
\langle {\cal O} \rangle 
= 
\frac{\int_{\cal C} \prod_x d\phi_x e^{-S[\phi]}{\cal O}[\phi]}
{\int_{\cal C} \prod_x d\phi_x e^{-S[\phi]}}
=
\frac{\sum_{\sigma}n_{\sigma}\int_{{\cal J}_{\sigma}} \prod_x d\phi_x e^{-S[\phi]}{\cal O}[\phi]}
{\sum_{\sigma}n_{\sigma}\int_{{\cal J}_{\sigma}} \prod_x d\phi_x e^{-S[\phi]}},
\]

but this is not realistic for many reasons: we have no systematic way to locate all the stationary points in a high
dimensional complex space, nor to compute the integer $n_{\sigma}$, nor to compute the {\em complex} factors
$Z_{\sigma}:=\int_{{\cal J}_{\sigma}} \prod_x d\phi_x e^{-S[\phi]}$ (that are also necessary to set up a Monte
Carlo calculation).

However, including all the thimbles ${\cal J}_{\sigma}$ corresponds to reproduce the original integral exactly.  We
actually have much more freedom to regularise a QFT, can we exploit it to justify a simplified expression?  In the
following section I examine two independent arguments along this line.

\section{Justification of the approach}

\subsection{The large volume point of view}
The first argument applies in the large volume (hence thermodynamic) limit, and it follows essentially
\cite{Witten:2010cx} (Sec.~3.2), where the role of a large parameter is played by $\lambda$ rather than the volume.

The argument goes as follows. The decomposition in the basis of thimbles states (in the homology sense):
\[
{\cal C} = \sum_{\sigma} n_{\sigma} {\cal J}_{\sigma}
\]
The sum over ${\sigma}$ is meant over all the stationary points $\varphi_{\sigma}$ in $\mathbb{C}^n$.  But, let
$\varphi_{\rm min}$ be the only global minimum of the action $S_R[\phi]$ in the original domain ${\cal C}$, and
let:
\[
s^{\rm min} = S_R[\varphi_{\rm min}] 
= 
\min_{\varphi\in {\cal C}} S_R[\varphi].
\]
Then Morse/Picard-Lefschetz theory tells us that only those stationary points $\varphi_{\sigma}$ can contribute
such that $S_R[\varphi_{\sigma}]>s^{\rm min}$ (being otherwise $n_{\sigma}=0$), and are sufficiently close to
${\cal C}$, and they are suppressed as:
\[
e^{-S_R[\varphi_{\sigma}]+s^{\rm min}}.
\]
This suggests that only the thimbles attached to the global minima in ${\cal C}$ actually dominate.  However, we
cannot exclude the possible presence of a large number of stationary points close to the global minima.  Of course,
if there are multiple thimbles with (almost) degenerate $S_R[\varphi_{\sigma}]$ their relative phase is crucial to
determine possible cancellations, as they do appear \cite{Dunne:2015eaa}.

\subsection{The universality point of view}
\label{sec:univ}
The second argument is independent from the first one and it is also inspired by an observation made in
\cite{Witten:2010cx} (Sec.~2.6)\footnote{``If we do not require that the generalized integral $Z_{H,{\cal C}}(k)$
  should agree with the original Chern-Simon integral $Z_H(k)$ when $k$ is an integer, we have much more freedom in
  the choice of the cycle ${\cal C}$. We can take ${\cal C}$ to be any integer linear combination of Lefschetz
  thimbles (...)''}.  It starts from the observation that there is much freedom in the choice of the details of the
regularization of a QFT, as long as it satisfies its essential properties.  In particular, if we can find a path
integral that defines a local QFT with the same interactions, the same degrees of freedom, the same symmetries and
symmetry representations, and also the same perturbative expansion as the original formulation, then we have no
reason to regard this new formulation as less legitimate than the original one.  From this point of view, it
becomes irrelevant which thimbles dominate the original path integral. What matters instead is which thimbles (or
combinations of thimbles) enjoy the necessary properties.  By universality, we expect two different formulations
that fulfill the same properties to give the same physical predictions, in the continuum limit.  Two legitimate
formulations that lead to different physical predictions may still be interpreted as different phases of the same
theory.  But, if any of these formulations needs to be excluded as unphysical, then we need to recognise that the
corresponding QFT is ambiguous: it is not simply defined in terms of elegant first principles, but also by some
more requirements that must be specified clearly even if they might be less elegant than we would like.

I now review to what extent the above properties are fulfilled in the case of interesting QFTs and for accessible
thimbles.  As discussed in \cite{Cristoforetti:2012su} for the Bose gas and QCD, each thimble has the right degrees
of freedom by construction, and each thimble ${\cal J}_{\sigma}$ enjoys all the symmetries (and symmetry
representations) of the action $S[\phi]$, that are also preserved by the stationary point $\varphi_{\sigma}$ to
which ${\cal J}_{\sigma}$ is attached.  In general, most stationary points are expected to have little or no
symmetry.  In these cases, only the sum over unbroken symmetries is meaningful.

Moreover, perturbation theory around a point $\varphi_{\sigma}$ on the thimble coincides with the usual
perturbation theory around the same configuration, as long as the latter is defined.  The extension of this
analysis to many other QFTs is straightforward.

Note that, in most interesting cases, the global minimum $\varphi_{\rm min}$ of the action $S_R$ in the original
domain is also a stationary point.  Hence it defines a thimble ${\cal J}_0$ and a path integral
\begin{equation}
\langle {\cal O} \rangle
=
\frac{\int_{{\cal J}_0} \prod_x d\phi_x e^{-S[\phi]}{\cal O}[\phi]}
{\int_{{\cal J}_0} \prod_x d\phi_x e^{-S[\phi]}},
\label{eq:PI0}
\end{equation}
which is a natural candidate for these theories, if it fulfills the above properties.  I refer to
\cite{Cristoforetti:2012su} for details.

One essential property is {\em locality}.  This issue was dismissed too quickly in \cite{Cristoforetti:2012su} (and
it is not mentioned in \cite{Witten:2010cx}), so I discuss it in more detail here.

The action $S[\phi]$ is clearly local, at any point $\phi$ in the complexified space, including any thimble.  But
the domain of integration in the functional integral (\ref{eq:PI0}) is not defined in terms of manifestly local
conditions, in general.  The matter is much simpler if we observe that the integral in Eq.~(\ref{eq:PI0}) is
constant for any domain of integration that belongs to the same homology class of ${\cal J}_0$.  So, we may
consider the properties of the homology class rather than those of the thimble ${\cal J}_0$, which is defined in
terms of non transparent differential equations.  The homology class, instead, is characterized only in terms of
asymptotic conditions at infinity.  In the case of a bosonic QFT with local interaction of maximal power, say
$n=4$, these asymptotic conditions take forms like:
\begin{equation}
\alpha_i < {\rm Arg}[\phi_x^2 \phi_{x+\mu} \phi_{x+\nu}] < \beta_i,\qquad\qquad i=1,\ldots,N_{\rm cond},
\label{eq:locbos}
\end{equation}
for some constants $\alpha_i$, $\beta_i$.  So, in these cases, the domain of integration can be expressed ---
equivalently --- in term of manifestly local conditions\footnote{An interesting insight in these issues can be
  gained if we observe \cite{Garcia:1996np} that any combination of thimbles solves the Schwinger-Dyson equations,
  but for different boundary conditions, which are defined by different asymptotic conditions.}.

The same argument goes through for any bosonic theory, but it fails in the fermionic case, because the zeros of the
determinant of the fermionic matrix $Q$ introduce new asymptotic conditions that have a different form (relevant
when the field approach the singularity):
\begin{equation}
\alpha <  {\rm Arg}[\log \det(Q)< \beta,
\label{eq:locfer}
\end{equation}
which are not manifestly local, because the fermionic determinant $\det(Q)$ is not a local function of the fields.
Actually, we know that the fermionic determinant is the result of an integration over fermionic fields with only
local interacions, and it does not introduce any non locality.  However, this simple argument does not go through
and the question remains open.  Of course, locality can be examined a posteriori, by looking at the behaviour of
correlators, but this is can be very difficult to interpret.

Finally, we should ask whether a {\em Hamiltonian formulation} exists for the path integral in Eq.~(\ref{eq:PI0}).
It turns out\footnote{I thank Carl Bender for pointing me to the relevant literature.} that this problem has been
studied in the context of PT theory, which is closely related to the formulation of QFTs in non-trivial thimbles.
A Hamiltonian formulation was established for Quantum Mechanical systems with PT symmetry
\cite{Bender:2002vv,Mostafazadeh:2002hb,Mostafazadeh:2003gz,Jones:2009br}.  An explicit extension to QCD is not
available yet, but it might be possible if the charge conjugation operator takes the role of parity, as suggested
in \cite{Meisinger:2013zfa}.

\section{A Monte Carlo algorithm on a Lefschetz thimble}
\label{sec:lang}

After motivating the approach, I need to review how to build an algorithm to sample the phase space on a thimble.
I consider a scalar field theory first.  What I want to compute is the following:
\begin{equation}
\langle {\cal O} \rangle
=
\frac{1}{Z_0}
\int_{{\cal J}_0} \prod_x d\phi_x e^{-S[\phi]}{\cal O}[\phi]
=
\frac{1}{Z_0} e^{-S_I}
\int_{{\cal J}_0} \prod_x d\phi_x e^{-S_R[\phi]}{\cal O}[\phi],
\label{eq:tolang}
\end{equation}
where the last step holds because $S_I$ is constant on the thimble \cite{Witten:2010cx,Witten:2010zr}.  This leads
to a functional integral with a {\em real} action $S_R$, which is bounded from below in ${\cal J}_0$, so we can use
any Monte Carlo method that is able to stay on the thimble ${\cal J}_0$.  In particular, we may consider a Langevin
algorithm based on the stochastic Langevin equations (including some internal index $a$):
\begin{eqnarray*}
\label{eq:SDLang}
\frac{d}{d\tau} \phi^{(R)}_{a,x} &=& -\frac{\delta S_R}{\delta \phi^{(R)}_{a,x}} + \eta^{(R)}_{a,x}  \\
\frac{d}{d\tau} \phi^{(I)}_{a,x} &=& -\frac{\delta S_R}{\delta \phi^{(I)}_{a,x}} + \eta^{(I)}_{a,x}.
\end{eqnarray*}
A great help comes from the fact that the drift term (first term above) preserves the thimble by construction,
because they are precisely the equations of SD (\ref{eq:SD}), expanded in terms of real and imaginary parts.  The
problem comes from the noise $\eta$, that needs to be projected on the tangent space of the thimble.  However, we
lack a local characterization of the thimble and computing its tangent space is not straightforward.

A solution appears if we add one more dimension to the system.  In fact, the tangent space at the stationary point
φ = φmin is usually easy to compute.  So, I can get tangent vectors at any point if I can transport a vector $\eta$
along the gradient flow $\partial S_R$, so that it remains tangent to ${\cal J}_0$.  This amounts to require that
the Lie derivative of the noise field along the gradient flow vanishes:
\[
{\cal L}_{\partial S_R} (\eta) = 0  \qquad \Leftrightarrow [\partial S_R, \eta] = 0;
\]
which is equivalent to a first order differential equation for the noise field:
\begin{equation}
\frac{d}{d\tau} \eta_{j}(\tau) = \sum_k \eta_{k}(\tau) \partial_{k}\partial_{j} S_R,
\label{eq:transport}
\end{equation}
which is easy to implement.

The Langevin algorithm is now possible by projecting the noise vector $\eta$ at the stationary point $\varphi_{\rm
  min}$, where the tangent space is known, and then transporting it according to Eq.~(\ref{eq:transport}) until the
current configuration where the noise needs to be applied.  Of course, each application of the noise leads to
departures from the thimble of $O(d\tau^2)$, that accumulate and need to be corrected.  This is done by ensuring
that the new configuration belong to the thimble, by following the equations of SD, and correcting.  Numerically,
this procedure can be made stable by formulating it as a $d+1$ dimensional boundary value problem.

\subsection{Residual phase}
\label{sec:resph}

As noticed at the beginning, the curves of SD define a curved manifold (the thimble) immersed in $\mathbb{C}^n$.
Hence, the measure term in Eq.~(\ref{eq:tolang}) is a determinant of a unitary matrix, which is a complex phase.
This {\em residual} phase must be included when computing the integral, and it can be done by reweighting.  Does it
bring back the sign problem?  We cannot exclude it in principle, but the following considerations reassure us.  The
residual phase tells how much the orientation of the thimble differs from the tangent space at the stationary
point. But this orientation is not expected to oscillate widely, instead, it should interpolate smoothly between the
directions of SD at $\varphi_{\rm min}$ and the asymptotic directions of convergence of the integral.  Moreover,
configurations where the average phase $\langle d\phi\rangle \ll 1$ are more and more suppressed like $e^{-S_R}$.
In other words, there is a natural strong correlation between phase and weight, which is precisely what is missing
in presence of a bad sign problem.  

All these observations are only qualitative, but they are also supported by the quantitative evidence of the tests
presented in \cite{Fujii:2013sra}, where the average phase is consistently found to be $>0.99$.

Even though the residual phase might not reintroduce the sign problem, it still needs to be computed.  An
efficient, stochastic algorithm was presented in \cite{Cristoforetti:2014gsa}, which is estimated to scale like $V
N_{\tau}^2 N_R$, where $V$ is the volume, $N_{\tau}$ the number of steps in the extra dimension (along the curves
of SD) and $N_R$ is the number of stochastic vectors needed for the estimate.

\subsection{Alternative algorithms}

Besides, the Langevin algorithm reviewed above, other algorithms have been proposed: an HMC algorithm
\cite{Fujii:2013sra}, that, in its most recent formulation, is essentially as expensive as the Langevin algorithm
described above; two Metropolis algorithms \cite{Mukherjee:2013aga,Alexandru:2015xva} that are simpler and faster,
but have the risk of poor acceptance for large systems; and also an alternative algorithm \cite{DiRenzo:2015foa}
that ensures a great control of the thimble, even in difficult situations, at the price of limited scalability, in
its current form.

\section{Test with a Bose gas}

The first test bed for the approach was a complex scalar field theory with $U(1)$ symmetry, $\lambda\phi^4$
interaction and a chemical potential $\mu$ coupled to the $U(1)$ charge $j_0$
\cite{Cristoforetti:2013wha,Fujii:2013sra}:
\[
S= \int d^4 x [|\partial \phi|^2 + (m^2-\mu^2)|\phi|^2 + \mu j_0 + \lambda |\phi|^4],
\qquad
j_{\nu} := \phi^* \overset{\leftrightarrow}{\partial_{\nu}}\phi.
\]
When $\mu\neq0$ the action is not real, $\mbox{Re}[e^{-S}]$ is not positive and we have a sign problem, which is
actually very strong.  In fact, with phase quenched, the average phase is essentially zero in the interesting
region even on lattices as small as $4^4$, which makes ordinary reweighting hopeless.

However, this model has been treated successfully both with Complex Langevin \cite{Aarts:2008wh} and with a dual
variables reformulation \cite{Gattringer:2012df,Gattringer:2012ap,Endres:2006xu}.  Hence, it represents an ideal
case to test the thimble approach.

To date, the most accurate analysis of the Bose Gas on a thimble was reported in \cite{Fujii:2013sra}.  They used
an HMC algorithm with constraints and studied only small lattices.  The version of the algorithm used there was
very expensive, but a more recent version is roughly as expensive as a Langevin algorithm described in
Sec.~\ref{sec:lang} (personal communication).

The results of \cite{Fujii:2013sra} show that it is possible to achieve, in this case, an excellent control of the
thimble with full agreement with the known results.  Moreover, the residual phase was shown to stay consistently
above $0.99$ in all their simulation points.

\subsection{Approximating the thimble}

An obvious question is how precisely do we need to stay on the thimble to keep the sign problem under control and
without introducing biases on the results?  As for the sign problem, a poor approximation of the thimble may
introduce more noise from a non vanishing $e^{iS_I}$, but this is expected to appear gradually.  As for biases, the
results remain correct as long as the integration domain belongs to the homology class of the thimble.  A change in
the homology class should be signaled by the appearance of divergences.

The crudest approximation possible consists in integrating on the tangent space of the thimble at the stationary
point.  This would be the exact thimble for a Gaussian action (hence ``Gaussian thimble'').  If the sign problem
comes from the quadratic part of the action this change actually eliminates most of the sign problem.  If the
system wants to stay on sufficiently smooth configurations, we may not feel so much the wrong asymptotic directions
of convergence (Note that this is not a Gaussian approximation of the action: we still use the full action for
importance sampling).

Somewhat surprisingly, this very crude approximation delivers results in excellent agreement with the known results
\cite{Cristoforetti:2013wha}, up to lattices of size $8^4$.  The algorithm is also very cheap because, compared to
the Langevin algorithm in the traditional phase space, it only requires an additional FFT application for each
Langevin step.  Hence, this technique seems ideal for a first exploration of new models.  More details can be found
in \cite{Cristoforetti:2013wha}.

\section{The case of gauge theories}

The application of the LT approach to gauge theories requires more sophisticated technology than the one described
above, in order to deal properly with the gauge symmetry.  However, the appropriate technology has been developed
since the first application of Morse theory to gauge theories in \cite{Bott54,Bott56}, and it was recently reviewed
in \cite{Witten:2010cx}.  These basis made the extension of the Langevin algorithm to gauge theories straightforward
\cite{Cristoforetti:2012su} (but see also \cite{DiRenzo:2015epd}.

\section{The case of fermionic theories}

More problematic is the extension to fermionic theories.  From the conceptual point of view I have already noticed
in Sec.~\ref{sec:univ} that the proof of locality of a single symmetric thimble fails.  From the algorithmic point
of view the algorithm is still applicable \cite{Cristoforetti:2012su}, but more expensive.

In the last year, much work was devoted to the study of theories with fermions in the LT approach.  In particular,
Mukherjee and Cristoforetti \cite{Mukherjee:2014hsa} considers the Hubbard model on a flat thimble; Kanazawa and
Tanizaki \cite{Kanazawa:2014qma} consider, analytically, the $0-$dimensional Gross-Neveu, Nambu-Jona-Lasinio and
Chen-Simon models; Tanizaki \cite{Tanizaki:2014tua} considers, analytically, a $0-$dimensional $O(2)$ toy model;
Tanizaki, Hidaka and Hayata \cite{Tanizaki:2015rda} consider a one-site fermionic model.

Here I want to add more comments on three very recent works that include numerical simulations on non-trivial
thimbles with fermionic actions.  Di Renzo and Eruzzi \cite{DiRenzo:2015foa,Eruzzi:2015ptd} study a chiral random
matrix theory \cite{Mollgaard:2013qra}, and show that a single thimble is able to reproduce the exact result also
where phase quenched and naive Complex Langevin fail.

Very recently, Fujii, Kamata and Kikukawa \cite{Fujii:2015rdd,Fujii:2015vha,Fujii:2015bua} have studied,
analytically and numerically, within the LT approach, the $0+1$ dimensional Thirring model at various coupling
$\beta$.  By employing a ``uniform field'' model, they can also estimate, analytically, the isolated contributions
from different thimbles.  In the large $L$ limit, they find good agreement between the exact results and the
results obtained in the dominant thimble for large $\beta$ and for small $\beta$ outside the crossover region.
Instead, they find a discrepancy for small $\beta$ in the crossover region, which is corrected by the inclusion of
one more thimble (more precisely, a pair of symmetrically related thimbles).  These results are also confirmed by
the independent study \cite{Alexandru:2015xva}, obtained with a different (Metropolis) algorithm.

Low dimensional systems are not strongly constrained by symmetries.  Hence it is interesting that only one or two
thimbles --- out of the many existing ones --- are sufficient here.

\section{Relations with complex Langevin}

There are close relationships between the LT approach and the complex Langevin (CL) approach.  In particular, the
stationary points are the same.  However, the fields equations are very different: in the CL case the fields are
allowed to explore, in principle, the whole $\mathbb{C}^n$ phase space.  In the LT case the fields are constrained
to the chosen thimble.  In practice, there seems to be relationships between the distributions effectively sampled
by the LT and CL approaches.  Such relations have been partially investigated in
\cite{Eruzzi:2014nea,Aarts:2014nxa,Aarts:2013fpa}, but they are not yet understood.

In the LT approach, one has to chose which thimbles to study, and justify the choice physically.  A very
interesting issue is to understand how the CL algorithm deals with the possible presence of many thimbles.  Does it
somehow find the original combination of thimbles (or, anyway, a physically valuable one), including their
possible relative phases?  A very recent paper \cite{Hayata:2015lzj} (but see also \cite{Pehlevan:2007eq}) connects
a failure of the CL approach to the way the CL algorithm effectively samples the thimbles.  It seems that thinking
in terms of thimbles helps also a better understanding of the CL approach.


\section{Conclusions}

I have reviewed a recent proposal to regularise QFTs with a sign problem on a Lefschetz thimble.  I see the LT
approachas a way to translates an old --- seemingly intractable --- problem (the sign problem) into many new hard
--- but probably much more tractable --- problems: dealing with 5D-algorithm, dealing with a residual phase,
ensuring or checking locality, justifying a choice of thimbles.

The choice of thimble(s) must be justified either in terms of dominant contributions or in terms of first
principles.  This latter point of view introduces a new perspectives on the question of ``What is a QFT''
\cite{Dunne:2015eaa}.

In general, the extra dimension that is necessary to represent a thimble might be very long, in order to capture
the complexity of a gauge theory.  This is certainly possible, but by the same argument one might argue that all
Monte Carlo simulations of gauge theories should be hopeless.  At least in one case of non-trivial QFT we showed
that the extra-dimension can be actually very short, and that even a cheap algorithm can reproduce the correct
results.  I am looking forward to new exciting applications.

A code for the Langevin algorithm on the thimble is available at
\url{https://bitbucket.org/ggscorzato/thimble-monte-carlo}

\acknowledgments

I am very grateful to the organizers of the Lattice 2015 conference for the invitation to present this review talk
and for full support.

\bibliographystyle{JHEP} 
\bibliography{../../density}{}

\end{document}